\documentclass[a4paper]{article}
\usepackage{amsfonts}

\newcommand{\E}{\mathbb{E}}
\newcommand{\hh}{\hat{H}}
\newcommand{\M}{{P_{\bar{\alpha}}}}
\renewcommand{\alph}{p}
\newcommand{\sk}{^{\rm SK}}
\newcommand{\vb}{^{\rm VB}}
\renewcommand{\o}{_{\rm o}}
\newcommand{\e}{_{\rm e}}
\newcommand{\h}{h}

\usepackage{epsfig}
\usepackage{amssymb}
\usepackage{amsmath}
\usepackage{amsthm}
\usepackage{latexsym}
\usepackage{graphicx}
\def\be{\begin{equation}}
\def\ee{\end{equation}}
\def\bc{\begin{center}}
\def\ec{\end{center}}
\def\abar{\bar{\alpha}}

\newtheorem{theorem}{Theorem}[section]

\begin{document}

\title{Notes on the polynomial identities in random overlap structures}
\author{Peter Sollich \footnote{King's College London, Department of Mathematics, Strand,
    London WC2R 2LS, United Kingdom}, \ Adriano Barra  \footnote{Dipartimento di Fisica, Sapienza Universit\`a di Roma, Piazzale Aldo Moro 5, 00185
    Roma, Italy} }
\date{Dec 5 2011}
\maketitle

\begin{abstract}
\noindent In these notes we review first in some detail the
concept of {\em random overlap structure} (ROSt) applied to fully
connected and diluted spin glasses. We then sketch how to write
down the general term of the expansion of the energy part from the
Boltzmann ROSt (for the Sherrington-Kirkpatrick model) and the
corresponding term from the RaMOSt, which is the diluted extension
suitable for the Viana-Bray model.
\newline
From the ROSt energy term, a set of polynomial identities (often known
as Aizenman-Contucci or AC relations)
is shown to hold rigorously at every order because of a recursive
structure of these polynomials that we prove. We show also, however, that
this set is smaller than the full set of AC identities that is already known.
Furthermore, when investigating the RaMOSt energy for the diluted
counterpart, at higher orders, combinations of such AC identities appear, ultimately suggesting a crucial role for the entropy in generating these constraints in spin glasses.
\end{abstract}

\section{Introduction}

The study of mean field spin glasses is very challenging from both
a physical \cite{MPV} and a mathematical \cite{T} point of view.
Concerning the latter, an increasing amount of work has, in recent
years, developed sophisticated mathematical techniques and used these
to confirm several scenarios from
theoretical physics (e.g.\
\cite{aldo}\cite{S70}\cite{boviernuovo}\cite{dotsenko1}\cite{dotsenko2}\cite{limterm}\cite{t1}).
\newline
Despite all the results that have been obtained by several techniques
that avoid the replica trick (e.g.\ cavity field \cite{g2,quadratic},
stochastic stability \cite{contucci,pierluz}, stochastic calculus
\cite{comets,pastur} and others \cite{ALR,mpv}) ---
including, of fundamental importance, the correctness of the
Parisi expression for the free energy \cite{broken}\cite{t4} --- the
question of its uniqueness is still a subject for debate. This brings
with it also the question of whether ultrametricity, with all its
peculiarities, necessarily holds \cite{MPV}.
Recently, fundamental progress has been made connecting ultrametricity
to polynomial identities \cite{arguin2,panchenko1,panchenko2}, mainly Ghirlanda Guerra relations (GG) \cite{gg}, 
highlighting the importance of polynomial identities
in the analysis of mean field spin glasses.
\newline
One of the key approaches in the field is
the powerful and physically profound concept of Random Overlap
Structures (henceforth ROSt) introduced by Aizenman, Sims
and Starr in \cite{ass}.
\newline
In this work we want to deepen our understanding of a certain kind of
polynomial identities, known as {\em Aizenman-Contucci} identities (AC)
\cite{ac}. These characterize, in a sense, the peculiar structure
of the spin glass phase( as there is a deep link between AC polynomials and GG identities) \cite{gg}\cite{parisiSS} within the framework of ROSt, both for fully connected
(Sherrington-Kirkpatrick, SK \cite{g2}\cite{sk}) and for diluted
(Viana-Bray, VB \cite{gt2}\cite{viana}) systems.
\newline
We  show how to systematically derive AC
relations from the energy contribution of the Boltzman Random Overlap
Structure, once a Hamiltonian is given. Interestingly, we
find that only a subset of the whole set of known identities can be obtained.
Furthermore, when looking at the diluted counterpart, where
a Parisi theory has not yet been fully achieved, we show that at high
orders of expansion, the AC-like relations come out but combined into larger
identities: it is not trivially possible to split them again to show that
they are zero separately. 
\newline
In section \ref{partone} we introduce the
general concept of Random Overlap Structures. Then, in section
\ref{parttwo} we show in general terms our technique for finding the
desired polynomial identities.
In section $4$ we apply the idea to the Boltzmann ROSt for the SK
model, while in section $5$ we test it on the Boltzmann RaMOSt for the
VB case. Section $6$ is left for discussion and closes the
paper.

\section{Random overlap structures}\label{partone}

In a nutshell, the ROSt generalizes the single spin cavity
approach \cite{barra1}\cite{g}\cite{g2} into one of {\em several
(and possibly many) added} spins. These are in contact with a
larger ``bath'' with its own interaction matrix. The ROSt allows
the properties of this bath, including the overlaps between
different states, to be specified in a very flexible manner by a
trial random structure which interacts with the original set of cavity
spins. This then permits one to represent the
pressure of the SK model as the infimum over a family of such
trial structures in a set of probability spaces.

The Parisi ROSt \cite{arguin}\cite{ruellegrem}, which has states
lying on an ultrametric tree \cite{derrida}\cite{grem}, has the
property of optimality with respect to this principle (i.e. it is
one way of realizing the infimum). It is thought to coincide with
the (conceptually much simpler) Boltzmann ROSt \cite{g} introduced by
Guerra, which was shown to share with the former the same optimality.


\subsection{Introducing the ROSt for the SK}\label{peter}

Let us start from a system of $M+N$ spins: we label the $N$ spins
$\sigma_1,\ldots,\sigma_N$ and think of them as cavity spins, and
denote the $M$ spins by $\tau_1,\ldots,\tau_M$ and think of them
as the environment (the thermal bath) for the cavity.

The size $M$ of the bath is now made large, at fixed $N$. An
important effect of taking this limit is that the fields acting on
the cavity spins are dominated by their interactions with the bath
rather than their interactions with each other. In the limit
$M\to\infty$ the cavity spins then become effectively
non-interacting with each other and live in uncorrelated fields
whose statistics are governed by those of the bath. We will now
detail this important motivation for the ROSt approach.

We define and decompose the Hamiltonian $ H_{M+N}(\sigma,\tau) $
of the overall $M+N$-spin system as
\begin{eqnarray}\label{pet1} && H_{M+N}(\sigma,\tau) = -\frac{1}{\sqrt{M+N}}\sum_{1 \leq k<l \leq M}J_{kl}\tau_k\tau_l \\ \nonumber
&& -\frac{1}{\sqrt{M+N}}\sum_{1 \leq k \leq M, 1 \leq i \leq
N}\tilde{J}_{ki}\tau_k\sigma_i -\frac{1}{\sqrt{M+N}}\sum_{1 \leq
i<j \leq N}\hat{J}_{ij}\sigma_i\sigma_j,
\end{eqnarray}
where the relevant interaction variables $J_{kl}$,
$\tilde{J}_{ki}$ and $\hat{J}_{ij}$ are all independent standard
Gaussian random variables. Now call the first term in the
Hamiltonian (\ref{pet1}) $H_{M+N}(\tau)$ and write the second one
as $\sum_{1 \leq i \leq N} \tilde{h}_i(\tau)\sigma_i$ with
$\tilde{h}_i(\tau) = -\frac{1}{\sqrt{M+N}}\sum_{1 \leq k \leq
M}\tilde{J}_{ki}\tau_k$ which for large $M$ becomes
\be\label{BROS1}\tilde{h}_i(\tau) = -\frac{1}{\sqrt{M}}\sum_{1
\leq k \leq M}\tilde{J}_{ki}\tau_k. \ee The third contribution in
(\ref{pet1}), which has the interactions among the cavity spins,
is a sum over only $N^2$ terms. This is at most
$O(N^2/\sqrt{(M+N)})$ and goes to zero for $M \rightarrow \infty$
as anticipated.
\newline
Similarly the Hamiltonian governing the $M$-spin bath can be
written as
$$
H_{M}(\tau)= H_{M+N}(\tau) -
(\frac{1}{M}-\frac{1}{M+N})^{1/2}\sum_{1 \leq k<l \leq
M}\hat{J}_{kl}\tau_k\tau_l
$$
Here the random interactions $\hat{J}_{kl}$ are independent from
all others (and not related to the $\hat{J}_{ij}$ above, the
latter referring to $\sigma$-$\sigma$ interactions). The above
decomposition of $H_M(\tau)$ can be understood by noting that it
gives for each bond strength a variance of $1/(M+N) + 1/M -
1/(M+N) = 1/M$ as it should be. For large $M$ we can then write
$$
H_M(\tau)= H_{M+N}(\tau) + \sqrt{\frac{N}{2}}\hat{H}(\tau), \ \
\hat{H}(\tau) = -\frac{\sqrt{2}}{M}\sum_{1 \leq k<l \leq
M}\hat{J}_{kl}\tau_k\tau_l.
$$
Putting both together gives for large $M$ at any fixed $N$ for the
difference between the log partition functions of the $M+N$ and
$M$-psin systems
\be\label{preros} \mathbb{E}\ln \frac{Z_{M+N}(\beta)}{Z_M(\beta)}=
\mathbb{E} \ln \frac{ \sum_{\sigma,\tau}e^{-\beta H_{M+N}(\tau) -
\beta \sum_{i} \tilde{h}_i(\tau)\sigma_i}}{\sum_{\tau}e^{-\beta
H_{M+N}(\tau)-\beta (N/2)^{1/2}\hat{H}(\tau)}}, \ee
where $\mathbb{E}$ represents the disorder average over
the couplings.
\newline
Now let us call
\be\label{BROS2} \xi(\tau) = e^{-\beta H_{M+N}(\tau)}, \ee
and symmetrize $\hat{H}(\tau)$ w.r.t.\ the ordering of $k$ and $l$
by defining i.i.d.\ unit Gaussian random variables $J_{kl}$ for
{\em all} pairs $(k,l)$ such that 
\be \hat{H}(\tau) =
-\frac{1}{%
%
%
M}\sum_{1 \leq k \neq l \leq
M}\hat{J}_{kl}\tau_k\tau_l. \ee We can also add the diagonal terms
and modify $\hat{H}(\tau)$ to \be\label{BROS3} \hat{H}(\tau) =
-\frac{1}{%
%
%
M}\sum_{1 \leq k,l \leq
M}\hat{J}_{kl}\tau_k\tau_l. \ee The resulting extra Gaussian
random contribution is $\tau$-independent and so pulls through the
sum over all $\tau$ and the logarithm in (\ref{preros}) to appear
linearly in the expectation over disorder, where it then vanishes.
Calling $P(\beta)$ the thermodynamic pressure, defined in terms of
free energy density $f(\beta)$ as $P(\beta)=-\beta f(\beta)$ and
using $\mathbb{E}\ln Z_{M+N}(\beta) = (M+N)P(\beta)$ for large $M$
gives finally \be P(\beta)=
\frac{1}{N}\mathbb{E}\ln\frac{Z_{M+N}(\beta)}{Z_M(\beta)}=
\frac{1}{N}\mathbb{E}\ln \frac{\sum_{\sigma,\tau}\xi(\tau)
e^{-\beta\sum_i
\tilde{h}_i(\tau)\sigma_i}}{\sum_{\tau}\xi(\tau)e^{-\beta(N/2)^{1/2}\hat{H}(\tau)}}.
\label{alpha_ROSt_motivation} \ee The $\tilde{h}_i(\tau)$ and
$\hat{H}(\tau)$ are all zero mean Gaussian random variables. The
two families of variables are uncorrelated with each other, while
within the families the covariances are
\begin{eqnarray}
\mathbb{E} \tilde{h}_i(\tau)\tilde{h}_j(\tau')  &=& \delta_{ij}
\frac{1}{M}\sum_{1 \leq k \leq M} \tau_k\tau'_{k} = \delta_{ij}q(\tau,\tau') \\
\mathbb{E} \hat{H}(\tau)\hat{H}(\tau')  &=& \frac{1}{M^2}\sum_{1
\leq k,l \leq M}\tau_k\tau_l\tau'_k\tau'_l = q^2(\tau,\tau').
\end{eqnarray}

A Random Overlap Structure or ROSt is a generalization of the
above structure which allows one to describe more generally (for
example in terms of a Parisi ultrametric tree) the states $\tau$
of the bath for the cavity spins $\sigma$. Similarly the overlaps
between these states are left unspecified, and hence denoted with
a tilde, as are the weights $\xi(\tau)$.

\bigskip

Let us then start by defining a \emph{Random Overlap Structure}
$\mathcal{R}$ as a triple $(\Sigma, \tilde{q}, \xi)$ where
\begin{itemize}
  \item $\Sigma$ is a discrete space;
  \item $\xi: \Sigma\rightarrow\mathbb{R}_+$
  is a system of random weights;
  \item $\tilde{q}:\Sigma^2\rightarrow[-1, 1]$
  is a symmetric \emph{Overlap Kernel}, with $\tilde{q}(\tau,\tau)=1$.
\end{itemize}
Now consider two families of independent centred Gaussian random
variables $\tilde{h}_{.}$ and $\hat{H}$, defined on $\Sigma\ni
\tau$, such that there are $N$ variables $\tilde{h}_i(\tau)$, for
each $\tau$ and
\begin{equation}\label{covarianze}
\mathbb{E}(\tilde{h}_{i}(\tau)\tilde{h}_{j}(\tau^{\prime}))
=\tilde{q}(\tau,\tau^{\prime})\delta_{ij}\ ,\
\mathbb{E}(\hat{H}(\tau)\hat{H}(\tau^{\prime}))=
\tilde{q}^{2}(\tau,\tau^{\prime}).
\end{equation}
Then the \emph{Generalized Trial Pressure} can be written as
\begin{equation}\label{gtf}
G_{N}(\mathcal{R})=\frac 1N\mathbb{E}\ln \frac{\sum_{\sigma,
\tau}\xi(\tau) \exp(-\beta
\sum_{i=1}^{N}\tilde{h}_{i}(\tau)\sigma_{i})}{\sum_{\tau}\xi(\tau)
\exp(-\beta ({N}/{2})^{1/2}\hat{H}(\tau))}\ .
\end{equation}

In the following two subsections we outline the properties of the
ROSt defined above, following the presentation in
\cite{ass},\cite{g}: we state the required theorems concerning the
optimality of the ROSt and we introduce the Boltzman ROSt,
referring the interested reader to the original papers for the
proofs.

\subsection{The Boltzmann ROSt}

Equations (\ref{BROS1},\ref{BROS2},\ref{BROS3}) define the
Boltzmann ROSt~\cite{g}, with one exception. In the Hamiltonian
$H_{M+N}(\tau)$ that defines the weights $\xi(\tau)$, and is given
by the first term on the r.h.s.\ of (\ref{pet1}), the normalizing
prefactor $1/\sqrt{M+N}$ is replaced by $1/\sqrt{M}$. This is
equivalent to replacing this Hamiltonian by $H_M(\tau)$, which
represents only the interactions within the bath.
\newline
On a superficial level, this change is necessary to comply with
the general definition of a ROSt: the weights $\xi(\tau)$ must not
depend on $N$. At first sight it looks dangerous, however for the
dominant states $\tau$, $H_{M+N}(\tau)$ is $O(M)$, and the change
of the prefactor by $\sqrt{(M+N)/M}=1+N/(2M)$ modifies the
Hamiltonian by a term of $O(N)$ that remains non-negligible even
for large $M$. Fortunately, the prefactor shift can alternatively
be regarded as a slight temperature shift to $\beta^* =
\beta\sqrt{(M+N)/M}$. Evaluating (\ref{alpha_ROSt_motivation}) at
this temperature and using $\beta^* H_{M+N}(\tau)=\beta H_M(\tau)$
gives
\be P(\beta\sqrt{M+N/M}) = \frac{1}{N}\mathbb{E} \ln
\frac{\sum_{\sigma,\tau}e^{-\beta H_M(\tau)} e^{-\beta
\sqrt{(M+N)/M}\sum_i\tilde{h}_i(\tau)\sigma_i}}
{\sum_{\tau}e^{-\beta H_M(\tau)}
e^{-\beta\sqrt{M+N/M}(N/2)^{1/2}\hat{H}(\tau)}}. \ee The two
exponents on the r.h.s.\ where the factor $\sqrt{(M+N)/M}$ now
appears do not grow with $M$ so in them one can replace
$\sqrt{(M+N)/M}$ by $1$ for large $M$. Similarly, as the
pressure is a continuous function of the inverse temperature, the
l.h.s.\ tends to $P(\beta)$ for large $M$. This shows that
(\ref{alpha_ROSt_motivation}) remains correct if we define the
weights as $\xi(\tau)=\exp(-\beta H_M(\tau))$, as claimed.
%



If we now call $\mathcal{R}_{B}(M)$ the Boltzmann ROSt we have
just defined, one can prove the following
\begin{theorem}[Reversed Bound]
\begin{equation*}
-\beta f(\beta)
\geq\lim_{N\rightarrow\infty}\liminf_{M\rightarrow\infty}
G_{N}(\mathcal{R}_{B}(M))
\end{equation*}
\end{theorem}
The idea of the proof, which we do not elaborate here, is to
compare $G_{N}(\mathcal{R}_{B}(M))$ with
$$
\frac1N\mathbb{E}\ln\frac{Z_{N+M}(\beta)}{Z_M(\beta)}\ .
$$
The following theorem states that the generalized trial pressure
provides an upper bound on the SK model pressure, i.e.\ a lower
bound for the free energy.
\begin{theorem}[Generalized Bound]
\label{b4}
\begin{equation*}
-\beta f_N(\beta) \leq 
\inf_{\mathcal{R}} G_N(\mathcal{R})\ .
\end{equation*}
\end{theorem}
From the two previous theorems one gets immediately the following
\begin{theorem}[Extended Variational Principle]
\begin{equation*}
\label{evp} -\beta f(\beta) =
\lim_{N\rightarrow\infty}\inf_{\mathcal{R}}G_N(\mathcal{R}) .
\end{equation*}
\end{theorem}
The theorem implies that it is sufficient to limit our trial
functions to those depending on trial overlaps, like those in the
ROSt space, and expressed as the difference between a cavity term
and an internal energy part, as in the numerator and denominator of
the generalized trial pressure $G_N$.

We will first decompose the generalized trial pressure
$G(\mathcal{R})$ from (\ref{gtf}), evaluated for the Boltzmann
ROSt $\mathcal{R}_B(M)$, into two parts. Seeing as the weights
$\xi(\tau)=\exp(-\beta H_M(\tau))$ of the bath configurations
$\tau$ are simply Boltzmann weights, we can introduce the notation
\[
\omega(F)=\frac{\sum_\tau F(\tau) \exp(-\beta
H_M(\tau))}{\sum_\tau \exp(-\beta H_M(\tau))}
\]
for the bath Boltzmann state. The generalized trial pressure can
then be written as
\be G_{N}(\mathcal{R}_B(M))=\frac 1N\mathbb{E}\ln
\omega\left(\sum_{\sigma} \exp(-\beta
\sum_{i=1}^{N}\tilde{h}_{i}\sigma_{i})\right) - \frac{1}{N}
\mathbb{E}\ln \omega\left( \exp(-\beta
\sqrt{\frac{N}{2}}\hat{H})\right). \label{gtf_decomposition} \ee
We will see that this decomposition mirrors exactly the one in
\cite{barra1}, and therefore call the second term the ``internal
energy term'' and the first the ``entropy term''. These names are
not quite precise but act as convenient shorthands. If $e(\beta)$
is the SK internal energy, then the internal energy term is in
fact $-(1/2)\beta e(\beta)$, while the entropy term contains in
addition to the entropy a contribution of $-(3/2)\beta e(\beta)$.

\bigskip

Let us now consider the internal energy term of the Boltzmann ROSt
generalized trial pressure, but with the $\beta$ in the exponent
generalized to some $\beta'$ that can be different from the
inverse temperature defining the bath Boltzmann state. Thanks to the stochastic stability of the Gibbs measure \cite{pierluz}, one can show that this affects the result only through a prefactor~\cite{barra2, barra4}, as stated in the following

\begin{theorem}[The energy expression]\label{dong} For $M\to\infty$,
\begin{equation}
\label{deng} \frac{1}{N}\mathbb{E}\ln
\omega\exp\left(-\beta^{\prime}\sqrt{\frac {N}{2}}
\hat{H}(\tau)\right)=\frac{\beta^{\prime 2}}{4}(1-\langle
\tilde{q}^2_{12}\rangle).
\end{equation}
On the r.h.s., $\langle\cdot\rangle=\mathbb{E}\Omega(\cdot)$ and
$\Omega$ is the replicated bath Boltzmann state $\omega$.
\end{theorem}


\subsection{Introducing the RaMOSt for the VB}

The RaMOSt plays a role analogous to that of the ROSt for the SK when
dealing with
diluted system such as the Viana-Bray model (VB) \cite{viana}. A
fundamental difference is
that we now need another real parameter $\alpha$ to take into
account the connectivity of the underlying random graph. We recall
here that the VB model is a spin model defined on a random graph, with
interactions only present on the edges of the graph.
\newline
As in the SK case, to motivate the RaMOSt, consider a cavity of
$N$ spins $\sigma_1,\ldots,\sigma_N$ in a ``bath'' of $M$ spins
$\tau_1,\ldots,\tau_M$, with $M\gg N$. The aim is to obtain the
pressure by considering the free energy increment when going from
$M$ to $M+N$ spins, $P(\beta,\abar)=(1/N)\mathbb{E}[\ln
Z_{M+N}(\beta,\abar) - Z_N(\beta,\abar)]$. Based on experience
with the SK model, where we had to allow for a slight temperature
shift to construct the Boltzmann ROSt, we  allowed here for a
shifted connectivity $\abar$ which should approach our desired
connectivity $\alpha$ when $M/N\to\infty$. To write the partition
function of the $(M+N)$-spin system, decompose the Hamiltonian as
in \cite{barra4}
\be -H_{M+N}(\sigma,\tau,\abar) =
\sum_{\nu=1}^{P_{\zeta}}J_{\nu}\tau_{i_{\nu}}\tau_{l_{\nu}}+
\sum_{\nu=1}^{P_{\tilde\zeta}}\tilde{J}_{\nu}\tau_{\tilde
i_{\nu}}\sigma_{\tilde j_{\nu}}+
\sum_{\nu=1}^{P_{\zeta'}}J'_{\nu}\sigma_{j'_{\nu}}\sigma_{k'_{\nu}}\
. \ee
All the $J$-variables here are i.i.d.\ interaction strengths,
distributed symmetrically about zero (e.g.\ binary values $\pm 1$ as
used in the following, or zero mean Gaussian variables; the precise
choice should be unimportant as mean field spin glasses are thought to
display universality \cite{wu}). The spin indices $i_\nu$ etc are
uniformly distributed
across $\{1,\ldots,M\}$ or $\{1,\ldots,N\}$ as appropriate. The upper
summation limits $P_{\zeta}$ etc.\ are Poisson random variables with
mean number of bonds in each ``sector'' given by
$$
\zeta=\abar\frac{M^{2}}{M+N}\ ,\ \tilde\zeta=\abar\frac{2MN}{M+N}\
,\ \zeta'=\abar\frac{N^{2}}{M+N}\ .
$$
For example, the total mean number of bonds is by definition
$(M+N)\abar$; there are $M^2$ ``$\tau\tau$''spin pairs out of a
total of $(M+N)^2$ and hence $\zeta=\abar(M+N)[M^2/(M+N)^2]=\abar
M^2/(M+N)$. To make the $\tau\tau$ part of the Hamiltonian
equivalent to an $M$-spin Hamiltonian with connectivity $\alpha$,
we need $\zeta=\alpha M$ and thus
\begin{equation}
\abar = \alpha \frac{M+N}{M}\ .
\end{equation}
As $\zeta'\to 0$ for $M\to\infty$ at fixed $N$, this term in
$H_{M+N}$ can be discarded with probability one: as in the SK
case, making the bath large enough allows us to neglect
interactions of the cavity spins. Summarizing so far, we have for
large $M$
\begin{eqnarray}
Z_{M+N}(\beta, \abar)&=& \sum_{\sigma,\tau} \exp\left( \beta
\sum_{\nu=1}^{P_{\alpha M}}J_{\nu}\tau_{i_{\nu}}\tau_{l_{\nu}}+
\beta\sum_{\nu=1}^{P_{2\alpha N}}\tilde J_{\nu}\tau_{\tilde
i_{\nu}} \sigma_{\tilde j_{\nu}} \right) \nonumber
\\
&=& \sum_{\sigma,\tau} \exp\left(-\beta H_M(\tau,\alpha) -
\beta \tilde{H}(\sigma,\tau,\alpha)\right) \label{VB:Z_M+N}
\end{eqnarray}
where
\begin{equation}
\tilde{H}(\sigma,\tau,\alpha) = -\sum_{\nu=1}^{P_{2\alpha N}}
\tilde{J}_\nu \tau_{\tilde i_\nu}\sigma_{\tilde j_\nu} =
\sum_{j=1}^{N}\tilde{h}_{j}(\tau)\sigma_{j}\ . \label{VB:Htilde}
\end{equation}
Here $\tilde{h}_{j}(\tau)$ is the cavity field acting on
$\sigma_{j}$ defined by
\begin{equation*}
\tilde{h}_{j}(\tau)=-\sum_{\nu=1}^{P_{2\alpha}}\tilde
J_{\nu}^{j}\tau_{\tilde i_{\nu}^{j}},
\end{equation*}
and the index $j$ of $\tilde J_{\nu}^{j}$ and $\tilde
i_{\nu}^{j}$ indicates independent copies of the corresponding
random variables. The first form of $\tilde H$ given in
(\ref{VB:Htilde}) is more useful for our calculations, while the
second one emphasizes the physics: as in the SK model, each cavity
spin $\sigma_j$ experiences a cavity field arising from its
interaction with the bath. In the VB case, this field is due to a
Poisson-distributed number (with mean $2\alpha$) of interactions
with randomly chosen spins $\tau_{\tilde i_{\nu}^{j}}$ from the
bath.

To write the partition function of the $M$-spin system with
connectivity $\abar$ in a similar form, we write its number of
bonds as $P_{\abar M}=P_{\alpha M} + P_{\alpha N}$:
\begin{eqnarray}
Z_M(\beta,\abar) &=& \sum_\tau \exp\left( \beta
\sum_{\nu=1}^{P_{\alpha M}}J_{\nu}\tau_{i_{\nu}}\tau_{l_{\nu}}+
\beta \sum_{\nu=1}^{P_{\alpha N}}\hat J_{\nu}\tau_{\hat
i_{\nu}}\tau_{\hat l_{\nu}} \right)\nonumber
\\
&=& \sum_\tau \exp\left( -\beta H_M(\tau,\alpha) - \beta\hat
H(\tau,\alpha) \right) \label{VB:ZM}
\end{eqnarray}
where
\begin{equation}
\hat{H}(\tau, \alpha)= -\sum_{\nu=1}^{P_{\alpha N}} \hat{J}_\nu
\tau_{\hat i_\nu}\tau_{\hat l_\nu}\ .
\label{hatH_first}
\end{equation}
Defining Boltzmann weights $\xi(\tau)=\exp(-\beta
H_M(\tau,\alpha))$, we can then write for large $M$
\begin{equation}
P(\beta,\alpha) =
\frac{1}{N}\mathbb{E}\ln\frac{Z_{M+N}(\beta,\abar)}{Z_M(\beta,\abar)}
= \frac{1}{N}\mathbb{E}\ln\frac{\sum_{\sigma,\tau} \xi(\tau)e^{-
\beta \tilde{H}(\sigma,\tau,\alpha)}} {\sum_\tau \xi(\tau)
e^{-\beta\hat H(\tau,\alpha)}}.
\end{equation}
By default the connectivity at which the pressure is found in this
way is $\abar$, but we have already exploited the fact that for
large $M$ this tends to $\alpha$.

In order to think at the above representation of the pressure (which so far
we have mainly tried to motivate, without being rigorous) as the
generalized trial pressure of a Random Multi-Overlap Structure
(RaMOSt), we need to show that the statistics of $\tilde{H}$ and
$\hat{H}$ can be expressed in terms of multi-overlaps of the bath
states $\tau$. To see this, note that the definitions
(\ref{VB:Htilde}) and (\ref{hatH_first}) of both
quantities can be written in terms of sums over i.i.d.\ variables:

\begin{eqnarray}
\label{def:hatH}\hat{H}(\tau,\alpha) &=&
\sum_{\nu=1}^{P_{\alpha N}}\hat{h}^{\nu}(\tau), \\
\label{def:tildeH} \tilde{H}(\sigma,\tau,\alpha) &=&
\sum_{\nu=1}^{P_{2\alpha N}} \tilde{h}^{\nu}(\tau)\sigma_{\tilde
j^{\nu}},
\end{eqnarray}
where $\hat{h}^{\nu}(\tau)$ and $\tilde{h}^{\nu}(\tau)$ are
independent copies of random variables $\hat{h}(\tau)$ and
$\tilde{h}(\tau)$. The latter form two independent families of
random variables indexed by $\tau$, whose probability
distributions have even moments
\begin{eqnarray}
\label{def:hatH_nu} \mathbb{E}[\hat{h}(\tau^1)\cdots
\hat{h}(\tau^{2n})] &=& (\mathbb{E}J^{2n})\,
\tilde{q}_{2n}^{2}(\tau^1,\ldots,\tau^{2n}), \\
\label{def:tildeH_nu}
\mathbb{E}[\tilde{h}(\tau^1)\cdots\tilde{h}(\tau^{2n})] &=&
(\mathbb{E}J^{2n})\, \tilde{q}_{2n}(\tau^1,\ldots,\tau^{2n}),
\end{eqnarray}
while the odd moments vanish. These properties follow from the
fact that in our construction so far $\hat{h}(\tau)=\hat
J\tau_{\hat i}\tau_{\hat j}$ and $\tilde{h}(\tau)=\tilde
J\tau_{\tilde i}$. The bath multi-overlaps occurring above are
then
\begin{equation}
\tilde{q}_{2n}(\tau^1,\ldots,\tau^{2n})=\frac{1}{M}\sum_{1 \leq i
\leq M}\tau_i^{1}\cdots\tau_i^{2n}\ . \label{Boltzmann_RaMOSt}
\end{equation}

We can now generalize and allow generic ways of specifying the
states $\tau$ of the bath and their multi-overlaps $\tilde
q_{2n}$. At this point it is clear that we have outlined
essentially the same setting as the one we used for the SK model,
and the previous remarks allow us to introduce the \emph{Random
Multi-Overlap Structure} $\mathcal{R}$ as a triple $(\Sigma,
\{\tilde{q}_{2n}\}, \xi)$ where
\begin{itemize}
  \item $\Sigma$ is a discrete space;

  \item $\xi: \Sigma\rightarrow\mathbb{R}_+$
  is a system of random weights;

  \item $\tilde{q}_{2n}:\Sigma^{2n}\rightarrow[-1, 1] , n\in\mathbb{N}$
  is a \emph{Multi-Overlap Kernel}. This needs to be such that
{\rm (\ref{def:hatH_nu},\,\ref{def:tildeH_nu})} define valid
random variables $\hat{h}(\tau)$ and $\tilde{h}(\tau)$, and in
particular each $\tilde q_{2n}$ must be symmetric in its
arguments. The multi-overlap kernels for different $n$ must also
be linked by the following \emph{reduction property}: $\tilde
q_{2n+2}(\tau,\tau,\ldots)=\tilde q_{2n}(\ldots)$ for
$n\in\mathbb{N}$ and $\tilde q_2(\tau,\tau)=1$.

\end{itemize}
The generalized trial pressure for such a RaMOSt is then defined
as
\begin{equation}
G_N(\mathcal{R}) = \frac{\sum_{\sigma,\tau} \xi(\tau)e^{- \beta
\tilde{H}(\sigma,\tau,\alpha)}} {\sum_\tau \xi(\tau) e^{-\beta\hat
H(\tau,\alpha)}}, \label{VB:gtf}
\end{equation}
where the statistics of the random variables $\tilde
H(\sigma,\tau,\alpha)$ and $\hat H(\tau,\alpha)$ are as defined by
{\rm (\ref{def:hatH}--\ref{def:tildeH_nu})}.

Note that the factorization of
(\ref{def:hatH_nu},\ \ref{def:tildeH_nu}) implies that
$\hat{h}^\nu(\tau)$ and $\tilde{h}^\nu(\tau)$ can be written as
$\hat J\hat\epsilon^\nu(\tau)$ and $\tilde
J\tilde\epsilon^\nu(\tau)$. The reduction property of the kernel
then further shows that $\hat\epsilon(\tau)$ and
$\tilde\epsilon(\tau)$ are binary ($\pm 1$), because all their
even moments are unity
($\mathbb{E}[\hat{\epsilon}^{2n}(\tau)]=\tilde
q_{2n}^2(\tau,\ldots,\tau)=1$ and similarly for $\tilde\epsilon$)
while the odd ones vanish.

We will call the RaMOSt introduced above, where
$\Sigma=\{-1,+1\}^M$, $\xi(\tau)=\exp(-\beta
H_M(\tau,\alpha))$ are Boltzmann weights and the multi-overlaps
are as in (\ref{Boltzmann_RaMOSt}), the \emph{Boltzmann RaMOSt}
$\mathcal{R}_B(M)$. The reduction property is then entirely
natural: even numbers of replicas cancel to give e.g.
$q_4(\tau^1,\tau^1,\tau^1,\tau^2)=q_2(\tau^1,\tau^2)$ and
$q_4(\tau^1,\tau^1,\tau^2,\tau^2)= q_2(\tau^2,\tau^2)=1$.

The generality of the RaMOSt allows one, on the other hand, to
take $\Sigma$ (which is not necessarily $\{-1, +1\}^{M}$) as the
set of indices $\tau$ of the weights $\xi(\tau)$ constructed by
means of Random Probability Cascades of Poisson-Dirichlet
processes (see e.g.\ Ref.\ \cite{t1}). These cascades give rise to
nested chains of expectations of Parisi type, and reproduce the
Parisi Replica Symmetry Breaking theory if one interpolates
according to the iterative approach of Refs.~\cite{franz, t1}.

\subsection{The Boltzmann RaMOSt}\label{bounds}

Now to acquire familiarity with the RaMOSt framework we state
a package of theorems mirroring the Aizenman, Sims and Starr
theory for the SK free energy \cite{lds1}.
\newline
Consider for $t\in [0, 1]$ and a given RaMOSt $\mathcal{R}$ the
following \emph{interpolating Hamiltonian}
\begin{equation*}
H(\sigma,\tau,t)=H_N(\sigma,t\alpha)+\hat{H}(\tau,t\alpha)+\tilde{H}(\sigma,\tau,(1-t)\alpha)
\end{equation*}
and using the RaMOSt weights $\xi(\tau)$ define
\begin{equation}
\label{r} g_N(t)= \frac{1}{N}\mathbb{E}\ln
\frac{\sum_{\sigma,\tau} \xi(\tau)\exp(-\beta H(\sigma,\tau,t))}
{\sum_{\tau}\xi(\tau)\exp(-\beta \hat H(\tau,\alpha))}\ .
\end{equation}
Clearly then
\begin{eqnarray*}
 &&g_N(0) = G_N(\mathcal{R})\ , \\
 &&g_N(1) = -\beta f_N(\beta,\alpha)\ .
\end{eqnarray*}
Within this construction the following results easily follow \cite{ass,barra2}.
\begin{theorem}[Generalized Bound]
\label{b}
\begin{equation*}
P_N(\beta,\alpha)\equiv -\beta f_N(\beta,\alpha)\leq
\inf_{\mathcal{R}} G_N(\mathcal{R})\ . \label{rmostbound}
\end{equation*}
\end{theorem}
\begin{theorem}[Reversed Bound] 
\label{rb}
\begin{equation*}
P(\beta,\alpha)\equiv -\beta f(\beta,\alpha) \geq
\inf_{\mathcal{R}}G_N(\mathcal{R}) \ .
\end{equation*}
\end{theorem}
\begin{theorem}[Extended Variational Principle]
\label{VB:evp_theorem}
\begin{equation*}
\label{VB:evp} P(\beta,\alpha)\equiv -\beta f (\beta,\alpha)=
\lim_{N\rightarrow\infty}\inf_{\mathcal{R}}G_N(\mathcal{R}) \ .
\end{equation*}
\end{theorem}
\begin{theorem}[The energy expression]\label{lema}
Let $\omega$, $\langle\cdot\rangle$ be the usual Boltzmann-Gibbs
and quenched Boltzmann-Gibbs expectations at inverse temperature
$\beta$, associated with the Hamiltonian $H_{N}(\sigma, \alpha)$.
Then
\begin{equation}\label{stability}
\lim_{N\to\infty}\mathbb{E}\ln\omega\exp\bigg(\beta^{\prime}
\sum_{\nu=1}^{P_{\bar{\alpha}}}
J^{\prime}_{\nu}\sigma_{i^{\prime}_{\nu}}\sigma_{j^{\prime}_{\nu}}\bigg)
=\bar{\alpha}\sum_{n=1}^{\infty}\frac{1}{2n}
\tanh^{2n}(\beta^{\prime})(1-\langle q^{2}_{2n}\rangle)\ ,
\end{equation}
where the random variables
$\{J^{\prime}_{\nu}\}$, $\{i^{\prime}_{\nu}\}$ and $\{j^{\prime}_{\nu}\}$
are independent copies of the analogous random variables appearing
in the Hamiltonian in $\omega$, and $P_{\bar{\alpha}}$ is a Poisson
random variable with mean $\bar{\alpha}$. On the r.h.s.\ the quenched
Boltzmann-Gibbs expectation is of the square of the multi-overlap
$q_{2n}$ of $n$ replicas of the system, defined as in
(\ref{Boltzmann_RaMOSt}).
\end{theorem}

\section{The general expansion of the ``energy'' term}\label{parttwo}

We want to expand the ``energy'' term
\be
e = \E \ln \Omega(\exp[-\beta'\,\hh(\tau)])
\ee
in $\beta'$ in order to compare this expansion with the r.h.s.\ of Eq.\
(\ref{deng}) for the SK model and Eq.\ (\ref{stability}) for the VB
model. We use $\Omega$ to denote the Boltzmann measure of $\tau$
(whose form will not matter), both for a single replica and later also
for the corresponding replicated measure.
\newline
In this section, we find a suitably general form of the expansion
of $e$, which does not rely on the specific form of $\hh(\tau)$.

We expand first the exponential
\be
\Omega(\exp[-\beta'\hh(\tau)])=1+\sum_{n\geq 1}\frac{(-\beta')^n}{n!}
\Omega(\hh^n(\tau))
\ee
and then the log to get
\be
e = \sum_{m\geq 1} \frac{(-1)^{m-1}}{m} \sum_{n_1\ldots n_m\geq 1}
\frac{(-\beta')^{n_1+\ldots+n_m}}{n_1!\cdots n_m!}
\E \left[\Omega(\hh^{n_1}(\tau)) \cdots \Omega(\hh^{n_m}(\tau))\right].
\ee
The expectation appearing here can be rewritten as
\be
\E \Omega \left[\hh(\tau^1)\cdots \hh(\tau^1) \times \cdots \times
\hh(\tau^n)\cdots \hh(\tau^n)\right]
\ee
where replica $\tau^1$ appears $n_1$ times, $\tau^2$ appears $n_2$
times and so on. Now group terms according to $n=n_1+\ldots+n_m$,
bearing in mind that $n\geq m$, and use the shorthand
$\langle\ldots\rangle = \E\Omega(\ldots)$:
\begin{eqnarray} \nonumber
e &=& \sum_{n\geq 1} \frac{(-\beta')^n}{n!} \sum_{m=1}^n
\frac{(-1)^{m-1}}{m}  \sum_{n_1\ldots n_m\geq 1,n_1+\ldots+n_m=n}
\frac{n!}{n_1!\cdots n_m!}  \times  \\ &\times&
\langle\hh(\tau^1)\cdots \hh(\tau^1) \times \cdots \times
\hh(\tau^n)\cdots \hh(\tau^n)\rangle
\end{eqnarray}
The combinatorial factor ${n!}/(n_1!\cdots n_m!)$ just gives the
number of permutations of the replica indices inside the
$\langle\ldots\rangle$, so one can write equivalently
\be e = \sum_{n\geq 1} \frac{(-\beta')^n}{n!} \sum_{m=1}^n
\frac{(-1)^{m-1}}{m} \sum'_{1\leq a_1,\ldots,a_n\leq m}
\langle\hh(\tau^{a_1})\cdots \hh(\tau^{a_n})\rangle. \ee
The prime on the last sum corresponds to the constraints
$n_1,\ldots,n_m\geq1$: only terms in which each of the $m$ replicas
appears at least once are to be included. In other words, as an
identity between sets (where multiple occurrences count as one) we
must have $\{a_1,\ldots,a_n\}=\{1,\ldots,m\}$.

We now exploit permutation symmetry of replicas to modify the sum over
$a_1,\ldots,a_n$ by expanding its summation range. This looks more
complicated initially but will pay dividends shortly by producing
unrestricted sums. From permutation
symmetry, our expansion is unchanged if we let $a_1,\ldots,a_n$ take
values in some general subset $T$
of $\{1,\ldots,n\}$, of size $|T|=m$. The constraint on the
summation would then be $|\{a_1,\ldots,a_n\}|=T$. We can now sum over
all $n!/[m!(n-m)!]$ possible choices of $T$ and divide by this factor.
The possible assignments of $a_1,\ldots,a_n$ that result from this
summation over $T$ are clearly all distinct, and together give
precisely all the assignments of $a_1,\ldots,a_n$ --- in the now {\em
expanded} range $1,\ldots,n$ --- for which the set
$\{a_1,\ldots,a_n\}$ has exactly $m$ elements. If we denote this
constraint with a superscript $(m)$ on the sum, we have
\be
e = \sum_{n\geq 1} \frac{(-\beta')^n}{n!} \sum_{m=1}^n
\frac{(-1)^{m-1}}{m} \frac{m!(n-m)!}{n!}
\sum^{(m)}_{1\leq a_1,\ldots,a_n\leq n}
\langle\hh(\tau^{a_1})\cdots \hh(\tau^{a_n})\rangle,
\ee
but now the sum over $m$ together with the constrained sum over the
$a_1,\ldots,a_n$ just yields an unconstrained sum. We just need to
bear in mind that the coefficient is $m$-dependent, i.e.
\be
e = \sum_{n\geq 1} \frac{(-\beta')^n}{n!^2}
\sum_{1\leq a_1,\ldots,a_n\leq n} (-1)^{m-1} (m-1)!(n-m)!
\langle\hh(\tau^{a_1})\cdots \hh(\tau^{a_n})\rangle,
\label{e_general}
\ee
where now $m=|\{a_1,\ldots,a_n\}|$ is a function of the
$a_1,\ldots,a_n$ which counts the number of distinct members
in the set of replica indices $\{a_1,\ldots,a_n\}$. This is our
desired general expansion, where now only unconstrained sums
appear. In our cases of interest, the averages over $\hh$ vanish for
odd $n$ and we need only the even terms, i.e.\ after relabelling $n\to2n$
\be
e = \sum_{n\geq 1} \frac{\beta'^{2n}}{(2n)!^2}
\sum_{1\leq a_1,\ldots,a_{2n}\leq 2n} (-1)^{m-1} (m-1)!(2n-m)!
\langle\hh(\tau^{a_1})\cdots \hh(\tau^{a_{2n}})\rangle.
\label{e}
\ee

\section{Expansion in the SK model}

For the SK model, we want to verify whether the identities we get
are indeed of AC form \cite{ac}, and to determine if and which
subset of AC identities they produce. Note that the expansion
parameter in $e$ is in principle $\beta'\sqrt{N/2}$ instead of
$\beta'$. We keep $\beta'$ for now; the original version can be
retrieved at
any point trivially by reinstating $\beta'\to \beta'\sqrt{N/2}$. The
perturbation Hamiltonian in the exponent is
\be
\hh(\tau) = - \frac{1}{M}\sum_{kl} \hat{J}_{kl} \tau_k\tau_l
\ee
with the $\hat{J}_{kl}$ i.i.d.\ Gaussian random variables of zero mean
and unit variance.

To simplify the expansion (\ref{e}) we carry out part of the disorder
average, over the $\hat{J}_{kl}$. Consider
\begin{eqnarray}
\langle \hh(\tau^{a_1}) &\cdots& \hh(\tau^{a_{2n}})\rangle \ =
\\ \nonumber
&=& M^{-2n}\sum_{k_1,l_1,\ldots,k_{2n},l_{2n}} \E(\hat{J}_{k_1
l_1}\cdots \hat{J}_{k_{2n} l_{2n}})
\langle\tau^{a_1}_{k_1}\tau^{a_1}_{l_1}\cdots
\tau^{a_{2n}}_{k_{2n}}\tau^{a_{2n}}_{l_{2n}} \rangle,
\end{eqnarray}
Wick's theorem gives a sum over pairings of the various indices
$(k_1,l_1)$, $(k_2,l_2)$ etc., or equivalently pairings of the
replica indices $a_1,\ldots,a_{2n}$. This can be written as a sum over
permutations $\pi$ of $2n$ elements if we bear in mind that we then
overcount each pairing $2^n n!$ times: 
\begin{eqnarray}
\langle \hh(\tau^{a_1}) &\cdots& \hh(\tau^{a_{2n}})\rangle \ =
\nonumber
\\ \nonumber
&=& \frac{1}{2^n n!}
M^{-2n}\sum_\pi\sum_{k_1,l_1,\ldots,k_{n},l_{n}} \langle
\tau^{a_{\pi(1)}}_{k_1}\tau^{a_{\pi(1)}}_{l_1}
\tau^{a_{\pi(2)}}_{k_1}\tau^{a_{\pi(2)}}_{l_1} \cdots
\\ \nonumber && \cdots \tau^{a_{\pi(2n-1)}}_{k_{n}}\tau^{a_{\pi(2n-1)}}_{l_{n}}
\tau^{a_{\pi(2n)}}_{k_{n}}\tau^{a_{\pi(2n)}}_{l_{n}} \rangle =
\\ &=&
\frac{1}{2^n n!} \sum_\pi \langle
q^2_{a_{\pi(1)}a_{\pi(2)}}\cdots
q^2_{a_{\pi(2n-1)} a_{\pi(2n)}}\rangle.
\end{eqnarray}
Now we insert this into the general expansion (\ref{e}). Because the
summation over $a_1,\ldots,a_{2n}$ is symmetric, each permutation
$\pi$ gives the same contribution and the sum over $\pi$ therefore
just yields a factor $(2n)!$ so that
\be
e\sk = - \sum_{n\geq 1} \beta'{}^{2n} \langle E\sk_n\rangle
\label{eSK}
\ee
with
\be
E\sk_n = \frac{1}{2^n (2n)!n!}
\sum_{1\leq a_1,\ldots,a_{2n}\leq 2n} (-1)^m (m-1)!(2n-m)!
\langle q^2_{a_1 a_2}\cdots q^2_{a_{2n-1}a_{2n}}\rangle.
\label{ESK_unordered}
\ee

This form of the result is beginning to look useful, but there is the
complication that, when e.g.\ $a_1=a_2$, $q^2_{a_1a_2}=1$ so it looks
like various orders of $q$ are mixed. We therefore next show that the
sum can be restricted to the terms were $a_1\neq a_2$, $a_3\neq a_4$
etc. To see this, insert into the sum appearing in the expression for
$E\sk_n$ a factor
\begin{eqnarray}
1 &=&
[(1-\delta_{a_1a_2})+\delta_{a_1a_2}]\cdots
[(1-\delta_{a_{2n-1}a_{2n}})+\delta_{a_{2n-1}a_{2n}}]
\\ \nonumber
&=& (1-\delta_{a_1a_2})\cdots(1-\delta_{a_{2n-1}a_{2n}})
\\
&&{}+\delta_{a_1a_2}
(1-\delta_{a_3a_4})\cdots(1-\delta_{a_{2n-1}a_{2n}}) + \ldots
\nonumber
\\
&&{}+\delta_{a_1a_2}\delta_{a_3a_4}
(1-\delta_{a_5a_6})\cdots(1-\delta_{a_{2n-1}a_{2n}}) + \ldots
\nonumber
\\
&&{}+\ldots
\nonumber
\\
&&{}+\delta_{a_1a_2}\cdots\delta_{a_{2n-1}a_{2n}}. \nonumber
\end{eqnarray}
We want to show that all the terms containing at least one factor
$\delta_{ab}$ (i.e.\ all except those in the first line) vanish once
summed over. This is easy to see. Consider without loss of generality
$\delta_{a_{2n-1}a_{2n}}$ and fix all other summation indices.  Call
the set of these indices $S=\{a_1,\ldots,a_{2n-2}\}$ and its size
$s=|S|$.  Now do the summation over $a_{2n-1}=a_{2n}$ in (\ref{eSK}),
noting that the average $\langle\ldots\rangle$ is independent of which value
$a_{2n-1}$ takes (since $q^2_{a_{2n-1}a_{2n-1}}=1$). Only $m$ varies:
either $m=s$ if $a_{2n-1}\in S$, or $m=s+1$ if $a_{2n-1}\not\in
S$. There are $s$ values of $a_{2n-1}$ where the first case occurs,
and $2n-s$ values for the second case. Thus
\begin{eqnarray}\nonumber
&& \sum_{1\leq a_{2n-1}=a_{2n}\leq 2n} (-1)^{m} (m-1)!(2n-m)!
\langle
\ldots \rangle = \\
\nonumber && s(-1)^{s}(s-1)!(2n-s)!+(2n-s)(-1)^{s+1} s! (2n-s-1)!
= 0,
\end{eqnarray}
and any summation over the remaining indices $a_1,\ldots,a_{2n-2}$
(whether or not they contain further pairs of identical indices)
of course then also gives a vanishing result. There is one
exception to this argument: if $n=1$ then $s=0$, and $m=1$
whatever the value of $a_1=a_2$; here no cancelation can occur
(mathematically, the breakdown of the argument is reflected in the
appearance of the divergent factor $(s-1)!=(-1)!$ above). The
$n=1$ term is therefore separated off explicitly below.

We have now shown that in (\ref{eSK}) we need to consider only
distinct summation indices within each pair, i.e.\ $a_1\neq a_2$
etc. We can further order the indices within each pair and then
need to multiply by a factor $2^n$ to compensate, giving (the
$\delta_{n1}$ term accounts for the non-canceling term at $n=1$)
\begin{eqnarray}
E\sk_n &=& -\frac{1}{2}\delta_{n1}+\frac{1}{n!(2n)!}
\sum_{{\rm o.p.}\leq 2n} (-1)^{m} (m-1)!(2n-m)!\,
q^2_{a_1 a_2}\cdots q^2_{a_{2n-1}a_{2n}}
\label{E_SK}
\\
&=& -\frac{1}{2}\delta_{n1}+ \frac{1}{n!} \sum_{2\leq m\leq 2n}
\sum_{{\rm o.p.}\leq m}^{(m)} \frac{(-1)^{m}}{m}\,
q^2_{a_1 a_2}\cdots q^2_{a_{2n-1}a_{2n}}.
\label{E_SK_compressed}
\end{eqnarray}
The subscript ``o.p.'' indicates a sum over ordered pairs, $1\leq
a_1<a_2\leq 2n$, $1\leq a_3<a_4\leq 2n$ etc. In the last row of
eq. (\ref{E_SK_compressed})  we have re-introduced a sum over $m$
and a constrained sum over (ordered pairs of) replica indices with
$m$ distinct elements. We have then further compressed the
summation range of the replica indices to $1,\ldots,m$,
multiplying by $(2n)!/[m!(2n-m)!]$ to compensate. Notice that in
this last version, $E\sk_n$ is no longer symmetric under
permutation of the replicas. But as we only need $E\sk_n$ under
the expectation $\langle E\sk_n\rangle$, which is invariant to
permutations of replicas, this does not matter. In the same
manner, we will from now on treat expressions in terms of overlaps
as identical as long as they give the same expectation $\langle
\ldots \rangle$ (or, equivalently, without taking the expectation
but after symmetrizing over all permutations of the replicas).

We can now state the identities that follow from the ROSt energy expression
(\ref{deng}). This contains only terms of order $\beta'{}^2$ on the
r.h.s., so comparing with the expansion (\ref{eSK}) shows that
our desired identities are simply $\langle E_n\rangle =0$ for $n\geq
2$. We next obtain a simple
recursion for the $E_n$ which shows that all these identities are of
AC form as expected. Consider $n\geq 2$ and let $S$ and $s$ be as defined
as above from the first $2n-2$ summation indices, i.e.\
$S=\{a_1,\ldots,a_{2n-2}\}$ and $s=|S|$. We now start from
(\ref{E_SK}) and make a transformation similar to the one leading to
(\ref{E_SK_compressed}) but only for these first $2n-2$ summation
indices. To this end we introduce a sum over $s=2,\ldots,2n-2$ and a
corresponding sum over (ordered pairs of) $a_1,\ldots,a_{2n-2}$
constrained so that $S$ has $s$ distinct elements. Permutation
symmetry tells us that we can compress the range of
$a_1,\ldots,a_{2n-2}$ from $1,\ldots,2n$ to $1,\ldots,s$, if we
multiply by the number of subsets of size $s$, $(2n)!/[s!(2n-s)!]$. In
this manner we get, if we abbreviate also $a=a_{2n-1}$ and $b=a_{2n}$,
\begin{eqnarray}
\label{E_SK_n_first}
E\sk_n & = &\frac{1}{n!}  \sum_{s=2}^{2n-2} \frac{1}{s!(2n-s)!}
\times
\\ \nonumber
& & \times \sum^{(s)}_{{\rm o.p.}\leq s}
\sum_{1\leq a<b\leq 2n} (-1)^{m} (m-1)!(2n-m)! q^2_{a_1
a_2}\cdots q^2_{a_{2n-3}a_{2n-2}}q^2_{ab}. \label{eSK_almostAC}
\end{eqnarray}
Now we carry out the sum over $a$ and $b$. The total number of
different replica indices present, $m$, depends on the range in which
$a$ and $b$ lie: if both are $\leq s$, we
have $m=s$, if only one is $>s$, $m=s+1$, and if both are $>s$, $m=s+2$.
Abbreviating $Q=q^2_{a_1 a_2}\cdots q^2_{a_{2n-3}a_{2n-2}}$,
the last sum from the previous equation becomes
\begin{eqnarray}
&& {} \sum_{1\leq a<b\leq s} (-1)^{s} (s-1)!(2n-s)! \,Q q^2_{ab}
\nonumber\\
 &+&{} \ \sum_{s<b\leq 2n}\sum_{1\leq a\leq s} (-1)^{s+1}
s!(2n-s-1)! \,Q q^2_{ab}
\nonumber\\
 &+&{}  \sum_{s< a<b\leq 2n} (-1)^{s+2} (s+1)!(2n-s-2)! \,Q
q^2_{ab}
\end{eqnarray}
Exploiting permutation symmetry among replica indices in the range
$s+1,\ldots 2n$ --- given that $Q$ is a function only of replicas
$1,\ldots,s$ --- and gathering prefactors simplifies this further to
\be
(-1)^s (s-1)!(2n-s)!\,
Q\left[ \sum_{1\leq a<b\leq s} q^2_{ab}
- s\sum_{1\leq a\leq s} q^2_{a,s+1}
+ \frac{s(s+1)}{2} q^2_{s+1,s+2}\right].
\ee
Denote the ``AC factor'' in the square brackets by $A_s$. Given that
this expressions will only be used under the expectation
$\langle\ldots\rangle$, which effectively symmetrizes it over
permutations of replicas, one can
use any integer larger than $s$ in
defining this factor, and in particular one can replace $A_s$ by
$A_{2n-2}$.

Overall, by inserting into (\ref{E_SK_n_first}) we can write the
coefficient $E\sk_n$ as
\be
E\sk_n = \frac{1}{n!}
\sum_{s=2}^{2n-2} \frac{(-1)^{s}}{s}
\sum^{(s)}_{{\rm o.p.}\leq s}
q^2_{a_1 a_2}\cdots q^2_{a_{2n-3}a_{2n-2}} A_{2n-2}
\ee
and so by comparison with (\ref{E_SK_compressed}) we get the elegant recursion
\be
E\sk_n = \frac{1}{n}(E\sk_{n-1}+\frac{1}{2}\delta_{n-1,1})A_{2n-2}.
\ee
Starting from $E_1=\frac{1}{2}(q_{12}^2-1)$, this gives the explicit
factorization (for $n\geq 2$)
\be
E\sk_n = \frac{1}{2n!}q_{12}^2 A_2 A_4 \cdots A_{2n-2}
\ee
This shows clearly that the identities $\langle E\sk_n\rangle=0$ for
$n\geq 2$ are in fact all of AC type. Each such identity corresponds
to the stochastic stability of the polynomial $E\sk_{n-1}$ of the
order below. Note that because we have already used permutation
symmetry to rewrite the summation over $a_1,\ldots,a_{2n-2}$ in terms
of $E_{n-1}$, the fully symmetric forms of the AC factors have to be
maintained, e.g.\ in $A_2=q_{12}^2-2(q_{13}^2+q_{23}^2)+3q_{34}^2$ one
cannot replace $q_{23}^2$ by $q_{13}^2$.

The explicit form of $E\sk_n$ shows that only a subset of all AC
identities is found from the energy term expansion: one has only
one $E_n$ for each $n$ (whereas from $n=3$ upwards there are more
stochastically stable monomials that one can use in place of
$E_{n-1}$ to produce different AC identities), and functions of
odd order like $q_{12}q_{13}q_{23}$ are missing altogether.

\section{Expansion in the VB model}

In the VB model the perturbation Hamiltonian is
\be
\hh(\tau) = \sum_{\nu=1}^\M \hat{J}_\nu\tau_{i_\nu}\tau_{j_\nu},
\ee
where $\M$
is a Poisson variable of mean $\bar\alpha$, $i_\nu$ and
$j_\nu$ for each $\nu$ are distributed uniformly over
$\{1,\ldots,M\}$, and $\hat{J}_\nu$ for each $\nu$ is $\pm1$ with equal
probability. The expectation in the general expansion (\ref{e}) is then
\begin{eqnarray}
 &&\hspace*{-1.5cm}
\langle \hh(\tau^{a_1})\cdots \hh(\tau^{a_{2n}})\rangle \ =
\nonumber
\\
&=& \E_\M \sum_{1\leq \nu_1,\ldots,\nu_{2n}\leq \M}
\E(\hat{J}_{\nu_1}\cdots \hat{J}_{\nu_{2n}})
\langle\tau^{a_1}_{i_{\nu_1}}\tau^{a_1}_{j_{\nu_1}}\cdots
\tau^{a_{2n}}_{i_{\nu_{2n}}}\tau^{a_{2n}}_{j_{\nu_{2n}}}
\rangle.
\end{eqnarray}
The average over the $\hat{J}$ vanishes except when the
$\nu_1,\ldots,\nu_{2n}$ coincide in pairs or larger groups of even
size, in which case it equals unity. The different patterns of groups
that can occur are precisely the even integer partitions of $2n$,
i.e.\ the integer partitions of $n$ multiplied by two. We
characterize such a partition of $n$ by the number of times
$k_\alph$ each integer $\alph$ occurs, such that $n=\sum_\alph
\alph k_\alph$ (where the sum over $\alph$ runs, here and in the
following, from 1 to $n$). For $n=3$, for example, the three
different partitions are $(k_1,k_2,k_3)=(3,0,0)$, $(1,1,0)$ and
$(0,0,1)$, corresponding to $3=1+1+1$, $3=1+2$, $3=3$. These
correspond respectively (after multiplication by two) to there
being three pairs of distinct $\nu$'s, one pair and one group of
four, and one group of six ($\nu_1=\ldots=\nu_6$). For each
partition, there are $(2n)!/(\prod_\alph (2\alph)!^{k_\alph}
k_\alph!)$ possibilities
--- remember that the group sizes in the partition of $2n$ are
$2\alph$, not $\alph$ --- of assigning $\nu_1,\ldots,\nu_{2n}$ and
the corresponding replica indices $a_1,\ldots,a_{2n}$ to groups of the
relevant sizes. Finally, given that each partition contains
$g=\sum_\alph k_\alph$ different groups, there are
$\M(\M-1)\cdots(\M-g+1)=\M!/(\M-g)!$ ways of assigning a value of $\nu$ to
each group.
Putting everything together gives, if $\sum_{(k)}$ denotes a sum
over all distinct integer partitions $(k_1,\ldots,k_n)$ of $n$,
\begin{eqnarray}
\langle \hh(\tau^{a_1})&\cdots &\hh(\tau^{a_{2n}})\rangle \ =
\nonumber
\\ \nonumber
&=& \sum_{(k)} \frac{(2n)!}{\prod_\alph (2\alph)!^{k_\alph}k_\alph!}
\E_\M [\M!/(\M-g)!]
\langle\tau^{a_1}_{i_1}\tau^{a_1}_{j_1}\cdots
\tau^{a_{2n}}_{i_g}\tau^{a_{2n}}_{j_g}\rangle
\label{partitions1}
\\
&=& \sum_{(k)} \frac{(2n)!\bar\alpha^g}{\prod_\alph
(2\alph)!^{k_\alph}k_\alph!} \langle
q^2_{a_1a_2\ldots}\times\cdots \times q^2_{\ldots
a_{2n-1}a_{2n}}\rangle. \label{partitions2}
\end{eqnarray}
Here the subscripts in the overlaps are arranged in accordance with
the specific partition considered, e.g.\ for $n=3$ and $(k)=(3,0,0)$
-- corresponding to $6=2+2+2$ -- the overlap product is
$q^2_{a_1a_2}q^2_{a_3a_4}q^2_{a_5a_6}$ while for $(k)=(1,1,0)$,
which corresponds to $6=4+2$, it is
$q^2_{a_1a_2a_3a_4}q^2_{a_5a_6}$. In the line above, the subscripts
$(i_1,j_1)$ to $(i_g,j_g)$ are arranged similarly, e.g.\ for
$(k)=(1,1,0)$ the first four replicas (or more precisely replica
indices) have subscripts $(i_1,j_1)$ and the last two have subscripts
$(i_2,j_2)$. Note that in this way we have picked out one particular
assignment of replica indices to the groups of the partition, and
multiplied accordingly with the number $(2n)!/(\prod_\alph
(2\alph)!^{k_\alph} k_\alph!)$ of such assignments. This is on the
understanding that the quenched average we are considering is to be
used inside a symmetric sum over $a_1,\ldots,a_{2n}$ (to get the
correct expression for a single setting of these summation variables we
would need to symmetrize by averaging over all permutations
$a_{\pi(1)},\ldots,a_{\pi(2n)}$).

To proceed, one inserts (\ref{partitions2}) into the general expansion
(\ref{e}). For the terms with $n\geq 2$ one could, as in the SK case,
switch to sums over ordered pairs, but this is not as useful here as
it does not prevent reductions in the order of the overlaps, e.g.\ we
would still get $q_{1212}^2 = 1$. We therefore leave the sum
unrestricted and write
\be
e\vb = -\sum_{n\geq 1} \beta'{}^{2n} \sum_{g=1}^{n}
\bar\alpha^g \langle E\vb_{ng}\rangle,
\label{e_VB}
\ee
where the coefficient $E\vb_{ng}$ is a sum over all integer partitions $(k)$ of
$n$ with $g$ terms:
\begin{eqnarray}
E\vb_{ng} &=& \sum_{(k):g=\sum_\alpha \! k_\alpha}
\frac{1}{\prod_\alph k_\alph!}
E_{(k)}
\label{E_VB_sum}
\\
 \label{E_VB_unordered}
E_{(k)} &=& \frac{1}{(2n)!\prod_\alph (2\alph)!^{k_\alph}}
\sum_{1\leq a_1,\ldots,a_{2n}\leq 2n} (-1)^{m}\times \\ \nonumber
& \times & (m-1)!(2n-m)!\, q^2_{a_1a_2\ldots}\times\cdots \times
q^2_{\ldots a_{2n-1}a_{2n}}
\end{eqnarray}
%

Let us discuss briefly how the above expression would change if
we were considering Gaussian couplings
$\hat{J}_\nu$. Here, within a group of size $p$, one would have $\E[\hat
J_\nu^{2p}]=(2p)!/(p!2^p)$ instead of $=1$ for the binary case. The
only change for the case of Gaussian couplings is therefore that in
(\ref{E_VB_unordered}) the factor $(2p)!^{k_p}$ is replaced by
$(p!2^p)^{k_p}$. 

We briefly compare (\ref{E_VB_sum},\ref{E_VB_unordered}) with the
corresponding result (\ref{eSK})
for the SK model, which after reinstating $\beta'\to
\beta'(N/2)^{1/2}$ reads
\be
e\sk = -\sum_{n\geq 1} (N\beta'{}^2/2)^n\langle E\sk_n\rangle.
\label{e_SK_with_N}
\ee
In the VB case, to get the Boltzmann RaMOSt we need to set
$\bar\alpha=N\alpha$. Defining also $\tilde\beta^2=2\alpha\beta'{}^2$ as
the equivalent SK temperature, the VB result (\ref{e_VB}) is
\begin{eqnarray}\nonumber
e\vb &=& -\sum_{n\geq 1} (N\alpha\beta'{}^2)^n \sum_{g=1}^n
(N\alpha)^{g-n} \langle E\vb_{ng}\rangle
\\ &=&  -\sum_{n\geq 1} (N\tilde\beta^2/2)^n
\sum_{g=1}^n
(N\alpha)^{g-n} \langle E\vb_{ng}\rangle.
\label{e_VB_with_N}
\end{eqnarray}
We see that, as it should be, precisely the SK contribution
(\ref{e_SK_with_N}) survives in the limit $\alpha\to\infty$ taken at
fixed $\tilde\beta$: this is the contribution with the largest number
$g=n$ of groups, where $(k)=(n,0,\ldots,0)$ and the combinatorial
factors are $1/\prod_\alph k_\alph!=1/n!$ and $1/\prod_\alph
(2\alph)!^{k_\alph}=1/2^n$,
%
%
giving $E\vb_{nn}=E_{(n,0,\ldots,0)}/n!=E\sk_n$ by
comparison with (\ref{ESK_unordered}).

We can now write down the AC-like identities for the VB model that are
obtained from the energy term (\ref{stability}) of
the RaMOSt. The r.h.s.\ of this exact expression contains only linear terms
in $\bar\alpha$, so comparison with (\ref{e_VB}) shows that we must
have $\langle E\vb_{ng}\rangle=0$ for $g\geq 2$ (hence from $n\geq g$
also $n\geq 2$).
In spite of the mixing of various orders of
(multi-)overlaps, one can still get a recursive factorization of the
$E\vb_{(k)}$ as we now show. Start from (\ref{E_VB_unordered}), and
suppose the last overlap in the product is of order $2p$
(corresponding to a term of this size in an integer partition of
$2n$). We want to do the sum over $a_{2(n-p)+1},\ldots,a_{2n}$.  By
analogy with the SK case, denote by $s$ the size of the set
$\{a_1,\ldots,a_{2(n-p)}\}$. Introduce a sum over $s$, and a
corresponding constrained sum over $a_1,\ldots,a_{2(n-p)}$; in the
latter, compress the summation range to $1,\ldots,s$ and multiply by
$(2n)!/[s!(2n-s)!]$ to make up for this. We get in this way
\begin{eqnarray}\nonumber
E_{(k)} &=& \frac{1}{\prod_{\alph'} (2\alph')!^{k_{\alph'}}}
\sum_{s=1}^{2(n-p)}\frac{1}{s!(2n-s)!} \times \\ \nonumber
&\times& \sum_{1\leq a_1,\ldots,a_{2(n-p)}\leq s}^{(s)}
q^2_{a_1a_2\ldots}\times\cdots \times q^2_{\ldots
  a_{2(n-p)-1}a_{2(n-p)}}
\nonumber
\\
&\times& \sum_{1\leq a_{2(n-p)+1},\ldots,a_{2n}\leq 2n} (-1)^{m}
(m-1)!(2n-m)!\, q^2_{a_{2(n-p)+1}\ldots a_{2n}}.
\label{VB_recursion1}
\end{eqnarray}
We now focus on the sum in the last line; call it $\Sigma$. The
complications in evaluating this arise because whenever a replica index occurs
twice (or more) it cancels and we get a lower order overlap. So we
need to consider again integer partitions, now of $2p$, to tell us how
such identical indices group. Let
$(\kappa)=(\kappa_1,\ldots,\kappa_{2p})$ denote such a partition, with
$\sum_r r\kappa_r=2p$ (where $r=1,\ldots,2p$ here and below). If
replica indices occur in groups of identical values according to such
a partition, all the even groups cancel completely from $q$, and we
get a multi-overlap of order $\gamma\o=\sum_{r\,{\rm odd}}\kappa_r$
given by the number of odd groups. Define also $\gamma\e=\sum_{r\,{\rm
even}}\kappa_r$, the number of even groups, and
$\gamma=\gamma\o+\gamma\e$, the total number of groups. For every
partition, there are $(2p)!/(\prod_r r!^{\kappa_r} \kappa_r!)$ ways of
arranging the replica indices into groups of the given size.

Finally, we need to account for which actual replica index value in
the range $1,\ldots,2n$ is used for each group of identical indices.
All groups need to have distinct index values (since groups are {\em
defined} as subsets of the summation variables
$a_{2(n-p)+1},\ldots,a_{2n}$ having identical values). We split the
sum over all possible assignments of index values according to the
number of groups of odd size, $\h\o\leq \gamma\o$, having {\em high}
index values $>s$, and the number of even groups, $\h\e\leq \gamma\e$,
with such high index values. We need these quantities to determine the
overall number $m$ of distinct index values, given that the values
$\{1,\ldots,s\}$ occur already in the first line of
(\ref{VB_recursion1}): $m=s+\h\o+\h\e$. An explicit summation is
required only over the $\gamma\o-\h\o$ (low) index values
$b_1,\ldots,b_{\gamma\o-\h\o}$ of the odd-sized groups that are in the
range $1,\ldots,s$.  The other (high) index values of the odd groups
we can set to $s+1,\ldots,s+\h\o$ by replica permutation symmetry. The
index values of the even groups we never need explicitly because they
disappear from the overlap. We take the $b_1,\ldots,b_{\gamma\o-\h\o}$
as ordered and then just need to work out how many index value
assignments there are for a given setting of these indices:
$\gamma\o!/[\h\o!(\gamma\o-\h\o)!]$ ways of choosing which odd groups
have high index values, and similarly
$\gamma\e!/[\h\e!(\gamma\e-\h\e)!]$ ways for the even groups;
$(\gamma\o-\h\o)!$ ways of permuting the
$b_1,\ldots,b_{\gamma\o-\h\o}$ among the odd groups with low index
values; $(2n-s)!/(2n-s-\h\o)!$ ways of assigning index values to the
odd groups with high indices (the order matters); and similar factors
$[s-(\gamma\o-\h\o)]!/[s-(\gamma\o-\h\o)-(\gamma\e-\h\e)]!$ and
$(2n-s-\h\o)!/(2n-s-\h\o-\h\e)!$ for the number of ways of assigning
index values to the even groups with low and high indices,
respectively. Putting everything together, we have
\begin{eqnarray}\nonumber
\Sigma &=& \sum_{(\kappa)} \frac{(2p)!}{\prod_r r!^{\kappa_r} \kappa_r!}
\sum_{\h\o,\h\e}
(-1)^{s+\h\o+\h\e} (s+\h\o+\h\e-1)!(2n-s-\h\o-\h\e)!
\label{Sigma_combinatorics}
\\
&&\times
\frac{\gamma\o!}{\h\o!(\gamma\o-\h\o)!}
\frac{\gamma\e!}{\h\e!(\gamma\e-\h\e)!}
(\gamma\o-\h\o)!
\frac{(2n-s)!}{(2n-s-\h\o)!}
\nonumber
\\
&& \times
\frac{[s-(\gamma\o-\h\o)]!}{[s-(\gamma\o-\h\o)-(\gamma\e-\h\e)]!}
\frac{(2n-s-\h\o)!}{(2n-s-\h\o-\h\e)!} \nonumber \\
&& \times \sum_{1\leq b_1<\ldots<b_{\gamma\o-\h\o} \leq s}
q^2_{b_1\ldots b_{\gamma\o-\h\o},s+1,\ldots,s+\h\o} \nonumber
\\ \nonumber
&=& \sum_{(\kappa)} \frac{(2p)!(2n-s)!}{\prod_r r!^{\kappa_r} \kappa_r!}
\sum_{\h\o}(-1)^{s+\h\o} \frac{\gamma\o![s-(\gamma\o-\h\o)]!}{\h\o!}
\\
&&\times \sum_{\h\e}
\frac{\gamma\e!}{\h\e!(\gamma\e-\h\e)!}(-1)^{\h\e}
\frac{(s+\h\o+\h\e-1)!}{[s-(\gamma\o-\h\o)-(\gamma\e-\h\e)]!} \nonumber \\
&& \times \sum_{1\leq b_1<\ldots<b_{\gamma\o-\h\o} \leq s}
q^2_{b_1\ldots b_{\gamma\o-\h\o},s+1,\ldots,s+\h\o}
\label{VB_calc}
\end{eqnarray}
The sum over $h\e$ can now be done, using that for $a\geq b$
\be
\sum_{k=0}^n \frac{n!}{k!(n-k)!} (-1)^k \frac{(a+k)!}{(b+k)!}
 = (-1)^n\frac{(a-b)!a!}{(a-b-n)!(b+n)!}
\label{factorial_sum}
\ee
(Factorials are treated like the corresponding Gamma functions here,
i.e.\ when $a-b-n$ is negative, $(a-b-n)!$ is infinite and the result
vanishes.) In our case $k\equiv \h\e$ with $n=\gamma\e$, $a=s+\h\o-1$,
$b=s+\h\o-\gamma\o-\gamma\e$ so that
\begin{eqnarray}
\Sigma &=& \sum_{(\kappa)} \frac{(2p)!(2n-s)!}{\prod_r
r!^{\kappa_r} \kappa_r!}
\sum_{\h\o}(-1)^{s+\h\o} \frac{\gamma\o![s-(\gamma\o-\h\o)]!}{\h\o!}
\label{VB_aux}
\\ \nonumber
&&\times
(-1)^{\gamma\e}
\frac{(\gamma\o+\gamma\e-1)!(s+\h\o-1)!}{(\gamma\o-1)!(s+\h\o-\gamma\o)!}
\\ \nonumber
&& \times \sum_{1\leq b_1<\ldots<b_{\gamma\o-\h\o} \leq s}
q^2_{b_1\ldots b_{\gamma\o-\h\o},s+1,\ldots,s+\h\o} \nonumber
\\ \nonumber
&=&\sum_{(\kappa)} \frac{(2p)!(-1)^s (s-1)!(2n-s)!}{\prod_r
r!^{\kappa_r} \kappa_r!}
(-1)^{\gamma\e} \gamma\o(\gamma\o+\gamma\e-1)! \,A^{(\gamma\o)}_{s}
\label{VB_aux2}
\end{eqnarray}
with the higher-order AC factor (note $A^{(2)}_s\equiv A_s$)
\be
A^{(\gamma\o)}_s = \sum_{\h\o=0}^{\gamma\o}
(-1)^{\h\o} \frac{(s+\h\o-1)!}{\h\o!(s-1)!}
\sum_{1\leq b_1<\ldots<b_{\gamma\o-\h\o} \leq s}
q^2_{b_1\ldots b_{\gamma\o-\h\o},s+1,\ldots,s+\h\o}
\label{AC_general}
\ee
Regarding the summation range for $\h\o$ in this definition, note that
in (\ref{VB_aux}) the factorials give the restriction that
$s+\h\o-\gamma\o\geq 0$, hence $\h\o\geq \gamma\o-s$. This is
ensured if
in (\ref{AC_general}) we assign the sum the value zero when there is
no possible ordered assignment of the summation variables because
there are more than $s$ such variables. For
$\gamma\o=0$, on the other hand, $\h\o=0$ also and so there are {\em no}
summation variables in (\ref{AC_general}). In this case the sum can be
set to unity, but in fact because of the factor $\gamma\o$ in
(\ref{VB_aux2}) we never need to evaluate $A^{(0)}_s$.
\newline
%
%
The factor $(s-1)!$ in (\ref{VB_aux2}), which diverges for $s=0$,
signals that this case needs to be checked separately. Because
$s=|\{a_1,\ldots,a_{2(n-p)}\}|$ this case can occur only if $n=p$,
in which
case $s=0$ is the {\em only} possible value. The apparent divergence
in the $(s-1)!$ factors can be traced back to the factor
$(s+\h\o+\h\e-1)!$ in (\ref{VB_calc}), which looks divergent for
$\h\o=\h\e=0$. However, when $s=0$, we must have
$\h\o=\gamma\o$, $\h\e=\gamma\e$ and $\Sigma$ from (\ref{VB_calc})
simplifies to
\be
\Sigma_0 = (2p)! \sum_{(\kappa)} \frac{(2p)!}{\prod_r r!^{\kappa_r} \kappa_r!}
(-1)^{\gamma\o+\gamma\e} (\gamma\o+\gamma\e-1)!\, q^2_{1 \ldots \gamma\o}.
\label{Sigma0}
\ee
Because $\gamma\o+\gamma\e=g\geq 1$, the potentially offending
factorial does indeed stay finite.
Note that for this $s=0$ expression the $\gamma\o=0$ term does not vanish
automatically as was the case for $s\geq 1$.
%
%
The factor $(-1)^{\gamma\o}$ can be dropped
because $\gamma\o$ must be even (since $2p$ is). Note that
(\ref{Sigma0}) can be regarded as a special case of (\ref{VB_aux2}) if
in the latter one first cancels the factor $(s-1)!$ (inserted into
$A_s^{(\gamma\o)}$ for $s\geq 1$ to get a simple form), then sets
$s=0$ and $n=p$, and finally cancels $\gamma\o(\h\o-1)!/\h\o!=1$ because
$\h\o=\gamma\o$.

To summarize, the sum $\Sigma$ takes the form (\ref{VB_aux2}) when
$s\geq 1$, which is always the case for $p<n$, while for $s=0$ and
hence $n=p$ it is given by (\ref{Sigma0}). We now just need to replace
the last line of (\ref{VB_recursion1}) by this to find, for $p<n$,
\begin{eqnarray}
\label{VB_recursion_almost}
E_{(k)} &=& \frac{1}{\prod_{\alph'} (2\alph')!^{k_{\alph'}}}
\sum_{s=1}^{2(n-p)}\frac{(-1)^s}{s} \sum_{1\leq
a_1,\ldots,a_{2(n-p)}\leq s}^{(s)} q^2_{a_1a_2\ldots}\times \cdots
\times
\\ \nonumber
& & \times q^2_{\ldots a_{2(n-p)-1}a_{2(n-p)}} \times
\sum_{(\kappa)} \frac{(2p)!}{\prod_r r!^{\kappa_r} \kappa_r!}
(-1)^{\gamma\e}\gamma\o(\gamma\o+\gamma\e-1)! \,A^{(\gamma\o)}_{s}
\end{eqnarray}
while for $n=p$ and hence $(k)=(0,\ldots,0,1)$
\begin{eqnarray}
E_{(0,\ldots,0,1)}
&=& \sum_{(\kappa)} \frac{1}{\prod_r r!^{\kappa_r} \kappa_r!}\,
(-1)^{\gamma\e} (\gamma\o+\gamma\e-1)!\, q^2_{1 \ldots \gamma\o}.
\label{E_VB_0}
\end{eqnarray}
To make (\ref{VB_recursion_almost}) into a recursion just takes a few
more steps now. Call the sum over partitions in the second line,
without the factor $(2p)!$,
\be
B^{(2p)}_{s} = \sum_{(\kappa)} \frac{1}{\prod_r
r!^{\kappa_r} \kappa_r!}
(-1)^{\gamma\e}\gamma\o(\gamma\o+\gamma\e-1)! \,A^{(\gamma\o)}_{s}
\label{B_def}
\ee
One now shows that, as for the second order case $\gamma\o=2$ and
because of permutation symmetry among replicas with indices $>s$, also
in the higher order AC factors $A^{(\gamma\o)}_s$ one can replace $s$
by any larger integer and in particular by $2(n-p)$. The same replacement
can then be made in $B^{(2p)}_{s}$. In the first line of
(\ref{VB_recursion_almost}), one re-expands the summation range on the
$a_1,\ldots,a_{2(n-p)}$, but now only to $1,\ldots,2n-2p$ and
correspondingly divides by $(2n-2p)!/[s!(2n-2p-s)!]$. The sum over $s$
and the constraint of having $s$ distinct summation indices can then
be combined into an unconstrained sum, where
$s=|\{a_1,\ldots,a_{2(n-p)}\}|$:
\begin{eqnarray}
E_{(k)} &=& \frac{1}{(2n-2p)!} \frac{1}{\prod_{\alph'}
(2\alph')!^{k_{\alph'}-\delta_{p'p}}}  \nonumber
\\ \nonumber &&\times \sum_{1\leq a_1,\ldots,a_{2(n-p)}\leq 2n-2p}
(-1)^s(s-1)!(2n-2p-s)!
\\
&&\times q^2_{a_1a_2\ldots}\times\cdots \times q^2_{\ldots
a_{2(n-p)-1}a_{2(n-p)}}\, B^{(2p)}_{2n-2p}.
\end{eqnarray}
Comparison with (\ref{E_VB_unordered}) now shows that
there is again a simple recursion:
\be
E_{(k)} = E_{(\ldots,k_p-1,\ldots)} B^{(2p)}_{2n-2p}
\label{VB_recursion}
\ee
and starting from (\ref{E_VB_0}) every $E_{(k)}$ can be expressed in
factorized form. The main difference between the VB and SK cases is
that the factors entering at each step of the recursion are a mixture
of AC factors of different orders (from 2 to $2p$). Also the final
coefficient (\ref{E_VB_sum}) at a given order is a sum over a number
of factorized expressions, one for each even integer partition of $2n$
containing the specified number $g$ of groups.

To get explicit expressions for the lowest order coefficients
$E\vb_{ng}$ we just need the initial values from (\ref{E_VB_0}) and
the factors $B_s^{(2p)}$. We start with the former: for $p=1$, there are
only two different partitions of $2p=2$:
\be
\begin{array}{lllll}
(\kappa) & \gamma\o & \gamma\e & \prod_r r!^{\kappa_r} \kappa_r!
& (-1)^{\gamma\e} (\gamma\o+\gamma\e-1)! \\
(0,1) & 0 & 1 & 2 & -1 \\
(2,0) & 2 & 0 & 2 & 1
\end{array}
\ee
so
\be
E_{(1)} = \frac{1}{2}(-1 + q_{12}^2)
\ee
as expected because this polynomial should equal $E\sk_1$. For $p=2$
one gets similarly
\be
\begin{array}{lllll}
(\kappa) & \gamma\o & \gamma\e & \prod_r r!^{\kappa_r} \kappa_r!
& (-1)^{\gamma\e} (\gamma\o+\gamma\e-1)! \\
(0,0,0,1) & 0 & 1 & 24 & -1 \\
(0,2,0,0) & 0 & 2 & 8 &  1 \\
(1,0,1,0) & 2 & 0 & 6 &  1 \\
(2,1,0,0) & 2 & 1 & 4 & -2 \\
(4,0,0,0) & 4 & 0 & 24 &  6
\end{array}
\ee
and thus
\be
E_{(0,1)} = \left(-\frac{1}{24} + \frac{1}{8}\right)
+ \left(\frac{1}{6} - \frac{2}{4}\right) q_{12}^2 +
\frac{6}{24} q_{1234}^2
\,=\, \frac{1}{12}(1 - 4 q_{12}^2 + 3 q_{1234}^2).
\ee
For the first $B$-factor $B_s^{(2)}$ we again need the integer partitions of
$2p=2$:
\be
\begin{array}{lllll}
(\kappa) & \gamma\o & \gamma\e & \prod_r r!^{\kappa_r} \kappa_r!
& (-1)^{\gamma\e} \gamma\o (\gamma\o+\gamma\e-1)! \\
(0,1) & 0 & 1 & 2 & 0 \\
(2,0) & 2 & 0 & 2 & 2
\end{array}
\ee
so
\be
B_s^{(2)}=A_s^{(2)}.
\ee
The first nontrivial $B$-factor is the one for $p=2$, where we need
integer partitions of $4$:
\be
\begin{array}{lllll}
(\kappa) & \gamma\o & \gamma\e & \prod_r r!^{\kappa_r} \kappa_r!
& (-1)^{\gamma\e} \gamma\o(\gamma\o+\gamma\e-1)! \\
(0,0,0,1) & 0 & 1 & 24 &  0 \\
(0,2,0,0) & 0 & 2 &  8 &  0 \\
(1,0,1,0) & 2 & 0 &  6 &  2 \\
(2,1,0,0) & 2 & 1 &  4 & -4 \\
(4,0,0,0) & 4 & 0 & 24 & 24
\end{array}
\ee
so that
\be
B_s^{(4)}=\left(\frac{2}{6}-\frac{4}{4}\right) A_s^{(2)}+
\frac{24}{24} A_s^{(4)} = - \frac{2}{3} A_s^{(2)} + A_s^{(4)}.
\ee
One sees that the terms with the highest-order multi-overlap follow a
simple pattern: in $E_{(0,\ldots,0,1)}$, $q_{1\ldots 2p}^2$ has a
prefactor of $1/(2p)$, while in the expression for
$B_s^{(2p)}$, the AC polynomial $A_s^{(2p)}$ occurs with unit
coefficient. Both of these observations follow from the fact that for
$\gamma\o=2p$ there are no even groups ($\gamma\e=0$) and the
partition of $2p$ has to be $(\kappa)=(2p,0,\ldots,0)$. In
(\ref{E_VB_0}) one has then $ (-1)^{\gamma\e}
(\gamma\o+\gamma\e-1)!/\prod_r r!^{\kappa_r} \kappa_r!  =
(2p-1)!/(2p)!=1/(2p)$, while in (\ref{B_def}), $(-1)^{\gamma\e}
\gamma\o(\gamma\o+\gamma\e-1)!/\prod_r r!^{\kappa_r} \kappa_r!
= 2p(2p-1)!/(2p)!=1$.

Equipped with the above initial values, the recursion
(\ref{VB_recursion}) and the definition (\ref{E_VB_sum}) we can now
get the relevant VB coefficients up to $n=4$:
\begin{eqnarray}
E\vb_{11} &=& E_{(1)} \,=\, \frac{1}{2}(-1 + q_{12}^2)
\,=\, E\sk_1
\\
E\vb_{21} &=& E_{(0,1)} \,=\,
\frac{1}{12}(1 - 4 q_{12}^2 + 3 q_{1234}^2)
\\
E\vb_{22} &=& \frac{1}{2} E_{(2,0)} \,=\, \frac{1}{2} E_{(1)}B_2^{(2)} =
\frac{1}{4}(-1+q_{12}^2) A_2^{(2)} \,=\, E\sk_2
\\
E\vb_{31} &=&  E_{(0,0,1)} \,=\,\ldots + \frac{1}{6} q_{123456}^2
\\
E\vb_{32} &=& E_{(1,1,0)} \,=\, E_{(1)}
B_2^{(4)} = \frac{1}{6}(-1+q_{12}^2)(-2A_2^{(2)}+3A_2^{(4)})
\\
\mbox{or} &=& E_{(0,1)}
B_4^{(2)} \,=\, \frac{1}{12}(1 - 4 q_{12}^2 + 3 q_{1234}^2)A_4^{(2)}
\\
E\vb_{33} &=& \frac{1}{6} E_{(3,0,0)} \,=\, \frac{1}{6} E_{(2,0)}
B_4^{(2)} \,=\, \frac{1}{6} E_{(1)} B_2^{(2)}B_4^{(2)} \\ &=&
\frac{1}{12}(-1 + q_{12}^2)A_2^{(2)}A_4^{(2)} \,=\, E\sk_3
\\
E\vb_{41} &=& E_{(0,0,0,1)} \,=\, \ldots + \frac{1}{8} q_{12345678}^2\\
E\vb_{42} &=& E_{(1,0,1,0)} + \frac{1}{2} E_{(0,2,0,0)} \,=\,
E_{(0,0,1)}B_6^{(2)} + \frac{1}{2}E_{(0,1)}B_4^{(4)}
\\
&=& \left(\ldots+\frac{1}{6}q_{123456}^2\right) A_6^{(2)}
+ \frac{(1-4 q_{12}^2 + 3
q_{1234}^2)}{24}\left(-\frac{2}{3}A_4^{(2)}+A_4^{(4)}\right)
\\
E\vb_{43} &=& \frac{1}{2} E_{(2,1,0,0)} \,=\, \frac{1}{2}
E_{(1,1,0)}A_6^{(2)} = \frac{1}{2} E_{(0,1)}A_4^{(2)}A_6^{(2)}
\\
&=& \frac{1}{24}(1-4 q_{12}^2 + 3 q_{1234}^2)A_4^{(2)}A_6^{(2)}
\\
E\vb_{44} &=& \frac{1}{24} E_{(4,0,0,0)} = \frac{1}{48} (-1+q_{12}^2)
A_2^{(2)}A_4^{(2)}A_6^{(2)} = E\sk_{4}
\end{eqnarray}
Note that in all terms with $g\geq 2$, which are the ones we are
interested in because they give us the identities $\langle
E\vb_{ng}\rangle = 0$, the constant contribution in the factor
$E_{(0,\ldots,0,1)}$ can be dropped. E.g.\ in $E\vb_{33}$, we can
symmetrize in $(-1/12)A_2^{(2)} A_4^{(2)}$ across the replica indices
$1,2,3,4$ because $A_4^{(2)}$ is symmetric in these indices
anyway. But under this symmetrization $A_2^{(2)}$ vanishes. We have
already exploited this in writing $E\vb_{22}=E\sk_2$,
$E\vb_{33}=E\sk_3$ and $E\vb_{44}=E\sk_4$ above.

Now consider the various identities that result in detail:
$\langle E\vb_{22}\rangle=0$ is, after dropping the constant in the
first factor, the standard $4^{\rm th}$ order AC
relation $\langle q_{12}^2 A_2^{(2)}\rangle=0$ \cite{barra4}. Next,
$\langle E\vb_{33}\rangle=0$ is, after
dropping the $-1$, the same identity as for the SK model,
$\langle q_{12}^2 A_2^{(2)} A_r^{(2)}\rangle=0$.
The relation $\langle
E\vb_{32}\rangle=0$ reduces to $\langle q_{1234}^2
A_4^{(2)}\rangle=0$ (the constant can be dropped, and $\langle
q_{12}^2A_4^{(2)}\rangle=\langle q_{12}^2A_2^{(2)}\rangle=0$) as in
\cite{ac}\cite{gg}. In
the first form written down above one can similarly reduce
everything to $\langle q_{12}^2 A_2^{(4)}\rangle=0$, which must
--- and does indeed --- give an equivalent relation,
but looks superficially different because we
have broken the replica permutation symmetry in a different manner.
However, this simple pattern (of $\langle q_{1\ldots
2n}A_{2n}^{(2)}\rangle =0$) does not persist to higher orders, as
the $n=4$, $g=2$ term shows: this is the first one where one gets
a sum over several partititions. After dropping terms that are
zero because of lower order identities, $\langle
E\vb_{42}\rangle=0$ becomes
\be
\frac{1}{24}\langle 4 q_{123456}^2 A_6^{(2)} + 3 q_{1234}^2 A_4^{(4)}
\rangle = 0,
\ee
and the two parts cannot be separated, at least not provably so from
the energy term expansion considered here.

In $\langle E\vb_{43}\rangle$, finally, the $q_{12}^2$-term can be
dropped because
$$q_{12}^2A_4^{(2)}A_6^{(2)}=q_{12}^2A_2^{(2)}A_4^{(2)}$$ which has
vanishing expectation due to the identity from $E\vb_{33}$. So one
gets
\be
\langle q_{1234}^2 A_4^{(2)}A_6^{(2)} \rangle = 0
\ee
$\langle E\vb_{44}\rangle =0$, finally, gives the 8-th order SK identity
$\langle q_{12}^2 A_2^{(2)}A_4^{(2)}A_6^{(2)} \rangle = 0$.

%

\section{Outlook}

The work presented in this paper was motivated by recent progress
\cite{arguin2,panchenko1,panchenko2} in our understanding of relations
among ultrametricity \cite{MPV} and polynomial identities \cite{ac,gg}
in mean field spin glasses. We first reviewed the concept of random
overlap structures \cite{ass}, both for fully connected and for
diluted disordered mean field spin systems. Then, starting from an explicit
expression for the {\em energy} within this framework, we
compared this to an expansion closer to the ones
obtained by stochastic stability \cite{contucci}\cite{parisiSS} or
smooth cavity field \cite{barra1} methods. We analysed the
resulting linear set of overlap identities (which usually develop
in statistical mechanics of quenched disordered systems), referred
to as Aizenman-Contucci equations \cite{ac}.
\newline
We extended previous results \cite{barra2}\cite{barra4} both by
deriving an alternative and more rigorous  recursive approach for
the derivation of these identities and by showing that, at least when
considering
the energy term of the Boltzmann ROSt/RaMOSt, the identities
obtained from the low orders of our expansion are in perfect agreement with the
same relations obtained with e.g.\ the replica trick. Going to higher
orders in the expansion, on the other hand, we found that the
resulting identities are fewer in number than the identities
known to hold for the Sherrington-Kirkpatrick \cite{gg} or the
Viana-Bray \cite{franz}  models.
\newline
As the Parisi solution of the SK model (encoded in Ruelle's
GREM \cite{ruellegrem} within this framework and called Parisi ROSt)
is known to satisfy
the whole set of AC identities \cite{parisitala}, our work strongly suggests
that these further, missing relations must be associated with
the entropic contribution of the ROSt, on which we plan to report soon.

\section*{Acknowledgements}

The authors are grateful to Francesco Guerra for useful conversations.
\newline
AB acknowledges the FIRB grant RBFR08EKEV.

\end{document}